\documentclass[aps,showpacs,twocolumn,floatfix,bibnotes]{revtex4}

\usepackage{dcolumn}
\usepackage{color}
\usepackage{amsmath}

\usepackage[dvips]{graphicx}

\newcommand\pictc[5]{\begin{figure}
                       \centerline{
                       \includegraphics[width=#1\columnwidth]{#3}}
                   \protect\caption{\protect\label{fig:#4} #5}
                    \end{figure}            }
\newcommand\pict[4][1.]{\pictc{#1}{!tb}{#2}{#3}{#4}}
\newcommand\rpict[1]{\ref{fig:#1}}

\newcommand{\be}{\begin{equation}}
\newcommand{\ee}{\end{equation}}
\newcommand{\bea}{\begin{eqnarray}}
\newcommand{\eea}{\end{eqnarray}}

\newcounter{Fig}

\begin{document}

\begin{sloppy}

\title{Tunable transmission and harmonic generation in nonlinear metamaterials}

\author{Ilya V. Shadrivov$^1$, Alexander B. Kozyrev$^2$,
Daniel W. van der Weide$^2$, and Yuri S. Kivshar$^1$}

\affiliation{$^1$Nonlinear Physics Center, Research School of
Physical Sciences and Engineering, Australian National University,
Canberra ACT 0200, Australia\\
$^2$Department of Electrical and Computer Engineering, University of
Wisconsin-Madison, Madison WI 53706, USA}

\begin{abstract}
We study the properties of tunable nonlinear metamaterial
operating at microwave frequencies. We fabricate the nonlinear metamaterial
composed of double split-ring resonators and wires where a varactor diode
is introduced into each resonator so that the magnetic resonance can
be tuned dynamically by varying the input power. We show that at higher powers 
the transmission of the metamaterial becomes power-dependent,and we demonstrate experimentally  power-dependent transmission properties and selective generation of higher harmonics.
\end{abstract}

\maketitle

Engineered microstructured
metamaterials demonstrate many intriguing properties for the
propagation of electromagnetic waves such as negative refraction.
Such materials have been studied extensively during recent years~\cite{review}. Typically, such
metamaterials are fabricated as composite structures created by many
identical resonant scattering elements with the size much smaller
than the wavelength of the propagating electromagnetic waves; such
microstructured materials can be described in terms of macroscopic
quantities-- electric permittivity $\epsilon$ and magnetic
permeability $\mu$. By designing the individual unit cell
of metamaterials, one may construct composites with effective
properties not occurring in nature.

Split-ring resonators (SRRs) are the key building blocks for the
composite metamaterials, in particularly the materials having the
negative refractive index~\cite{exp}. Recent theoretical studies
have demonstrated how to dynamically tune or modulate the
electromagnetic properties of
metamaterials~\cite{our_prl,gorkunov_tuning,non_1,non_2,non_3,our_review} and the
fabrication of nonlinear SRRs has been demonstrated by placing a
varactor diode~\cite{our_oe} or a photosensitive
semiconductor~\cite{exp_smith} within the gap of the resonator. The
diode allows the SRR element to be tuned by an applied dc voltage or
by a high-power signal as was shown already in
experiment~\cite{our_oe,our_apl}. These recent advances open a way
for both fabrication and systematic study of {\em nonlinear tunable
metamaterials} which may change their properties such as the
transmission characteristics by varying the amplitude of the input
electromagnetic field.

It was shown theoretically that nonlinear metamaterials can
demonstrate many intriguing features such as unconventional
bistability~\cite{our_prl,bistability}, backward phase-matching and
harmonic generation~\cite{shg,zharov,shalaev}, as well as parametric
shielding of electromagnetic fields~\cite{shield}. Some of these
features have already been experimentally observed in nonlinear
left-handed transmission lines which are model systems allowing for
combining nonlinearity and anomalous
dispersion~\cite{kozyrev1,kozyrev2,kozyrev3}. Moreover, first results on harmonic generation in infrared were reported in Ref.~\cite{wegener_shg}. Importantly, in such
composite structures the microscopic electric fields can become much
higher than the macroscopic electric field carried by the
propagating electromagnetic waves. This provides a simple physical
mechanism for enhancing nonlinear effects in the resonant structure
with the left-handed properties. Moreover, a very attractive goal is
to create tunable metamaterial structures where the field intensity
changes the properties of a composite structure enhancing or
suppressing the wave transmission.

\pict{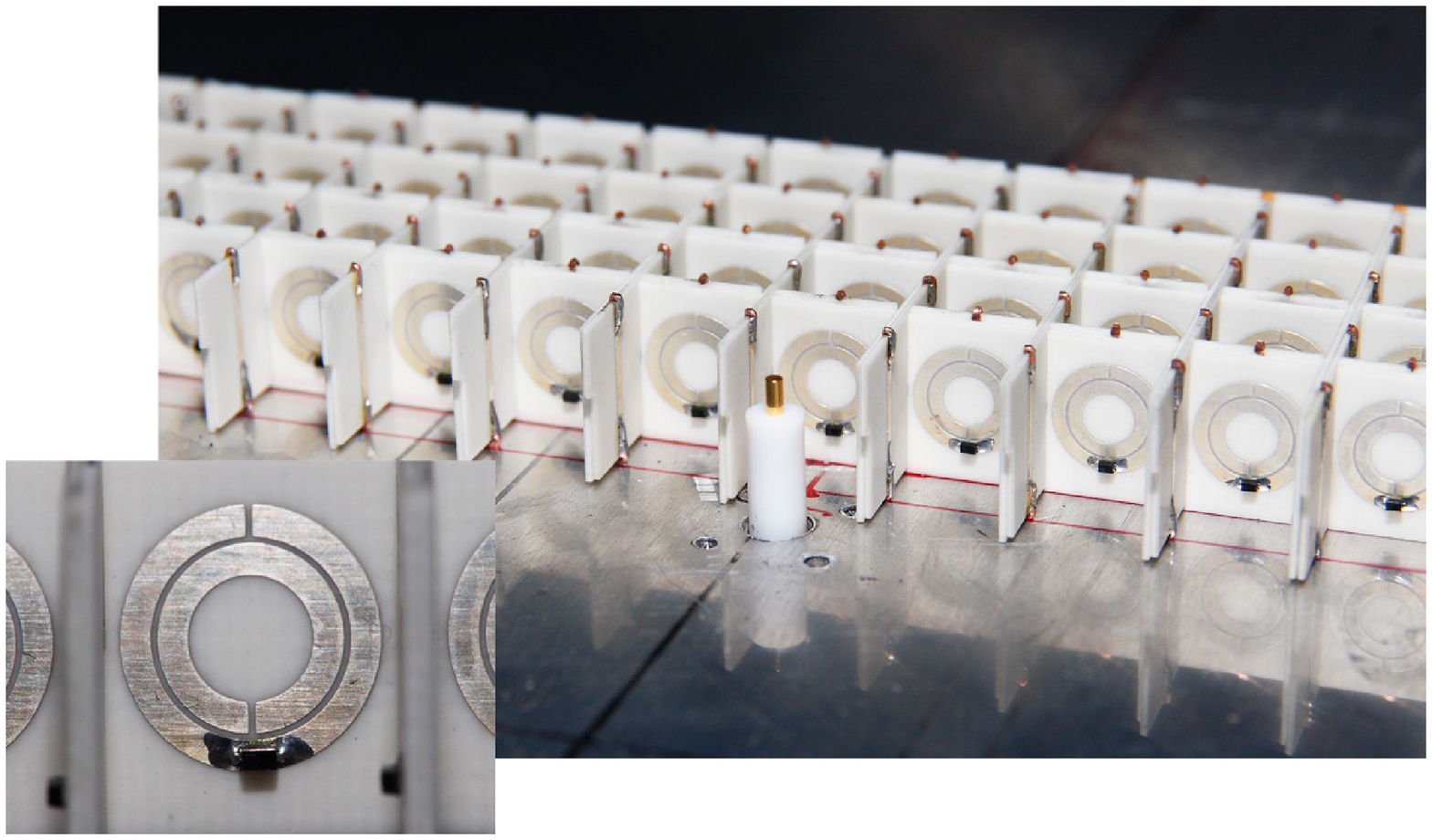}{photo}{(color online) Photograph of the nonlinear tunable metamaterial
created by square arrays of wires and nonlinear SRRs. Each SRR
contains a varactor (see also the enlarged photo in the inset) which
provides power-dependant nonlinear response.}

In this Letter, we report on the fabrication and experimental
studies of the properties of the first nonlinear tunable
metamaterial operating at microwave frequencies. Such a metamaterial
was fabricated by modifying the properties of SRRs and introducing
varactor diodes in each SRR element of the composite
structure~\cite{our_oe,our_apl}, such that the whole structure
becomes dynamically tunable by varying the amplitude of the
propagating electromagnetic waves. In particular, we demonstrate the
power-dependent transmission of the left-handed and magnetic
metamaterials at higher powers, as was suggested earlier
theoretically~\cite{our_prl} and selective generation of higher
harmonics.

 \pict{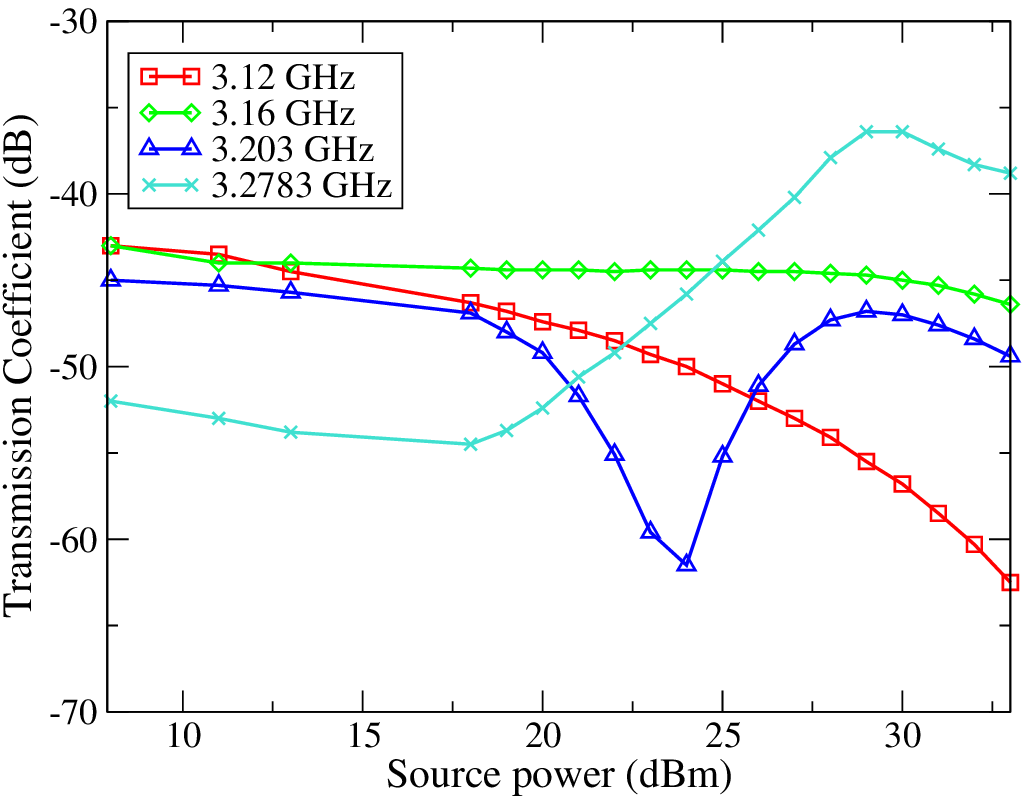}{Transmission_LHM}{(color online) Experimentally
 measured transmitted power (top) and transmission coefficient
 (bottom) vs. output power for different frequencies. Depending on
 the frequency, the metamaterial behaves as linear (green curve) or
 strongly nonlinear with different types of nonlinearity (three
 remaining curves).}

Metamaterial sample (see
Fig.~\rpict{photo}) is fabricated from 0.5 mm thick Rogers R4003
printed circuit boards with nominal dielectric constant of 3.4. Our
design allows for reconfiguring metamaterial structures by
rearranging the same boards. In particular, we make {\em three sets}
of dielectric boards with the appropriate slot allocations: (i)
without any metallization; (ii) with tin coated copper nonlinear
SRRs; (iii) with tin coated copper wires. Combining these boards in
different pairs, we can achieve different types of composite media,
including a wire medium; nonlinear magnetic metamaterial (MMM); and
nonlinear left-handed metamaterial (LHM). Photograph of one of the
nonlinear LHM structures is shown in Fig.~\rpict{photo}. Each SRR
contains variable capacity diode (model Skyworks SMV-1405) which
introduces nonlinear current-voltage dependence and resulting
nonlinear magnetic dipole moment to each SRR ~\cite{our_apl}. In
terms of effective medium parameters, the manufactured structure has
nonlinear magnetization and nonlinear effective magnetic
permittivity~\cite{our_prl}. Arrays of SRRs and wires form a square
lattice with 29x4x1 unit cells of the size of 10.5mm.

 First, to identify the
effect of the nonlinearity we measure the transmission properties of
this tunable metamaterial for different values of the input power.
To measure the electromagnetic field scattering on our samples, the
metamaterial slab is placed in a parallel plate waveguide. The
planes of SRRs are aligned perpendicular to the parallel plate
surfaces. The input antenna is placed at the midpoint of the lower
plate, 2~mm from the metamaterial slab, in front of the central unit
cell, and it consists of a teflon-coated conductor of 1.26~mm
diameter and 11~mm long. The teflon coating provides a better energy
coupling into the waveguide for the wavelengths of interest. The
antenna is positioned perpendicular to the bottom plate, so that the excited electric field is polarized perpendicular to the plane, and thus parallel to the wires. The magnetic field of the wave has mainly an in-plane component, effectively exciting the SRRs. Close
positioning of the source antenna to the metamaterial was chosen in
order to funnel high EM power into the metamaterial sample in
order to observe nonlinear effects. We note that different positioning of the source antenna with respect to the central unit cell of the metamaterial gives slightly different quantitative results for the measured transmission, however qualitatively all the results are identical. This effect appears due to different antenna impedance matching to the sample. An identical antenna is placed
in the center of the top plate, and is used as receiver for spectra
measurements and for raster scan of the electric field distribution
in the horizontal plane. The input antenna is excited using an
Agilent E8364A vector network analyzer, which output is amplified by
HP 83020A 38dB amplifier. For the transmission measurements, the receiving antenna is located 2~cm
behind the metamaterial slab, in front of the central unit cell of the metamaterial, and it is connected to the network analyzer as well. The measurements of the electric field inside the
waveguide are evaluated in terms of the magnitude and phase of the transmission coefficient
$S_{21}$ between the input of the source and output of the receiver antenna. Due to the
two-dimensional nature of the parallel plate waveguide, as well as
symmetry of our sample, the electric field in the scanned area is
expected to remain polarized mainly perpendicular to the plane of
the plates.

Figure~\rpict{Transmission_LHM} shows the experimentally
measured transmission coefficient as functions of the source power for different frequencies. The
measurement is performed for the SRR-wire structure. Due to the
nonlinearity-induced shift of the resonant frequency of SRRs,
studied earlier for a single SRR~\cite{our_apl}, the transmission of
the metamaterial sample becomes dependent on the input power. In
particular, depending on the value of the frequency, the
metamaterial behaves as linear or strongly nonlinear, as shown in
Fig.~\rpict{Transmission_LHM}.

\pict[0.95]{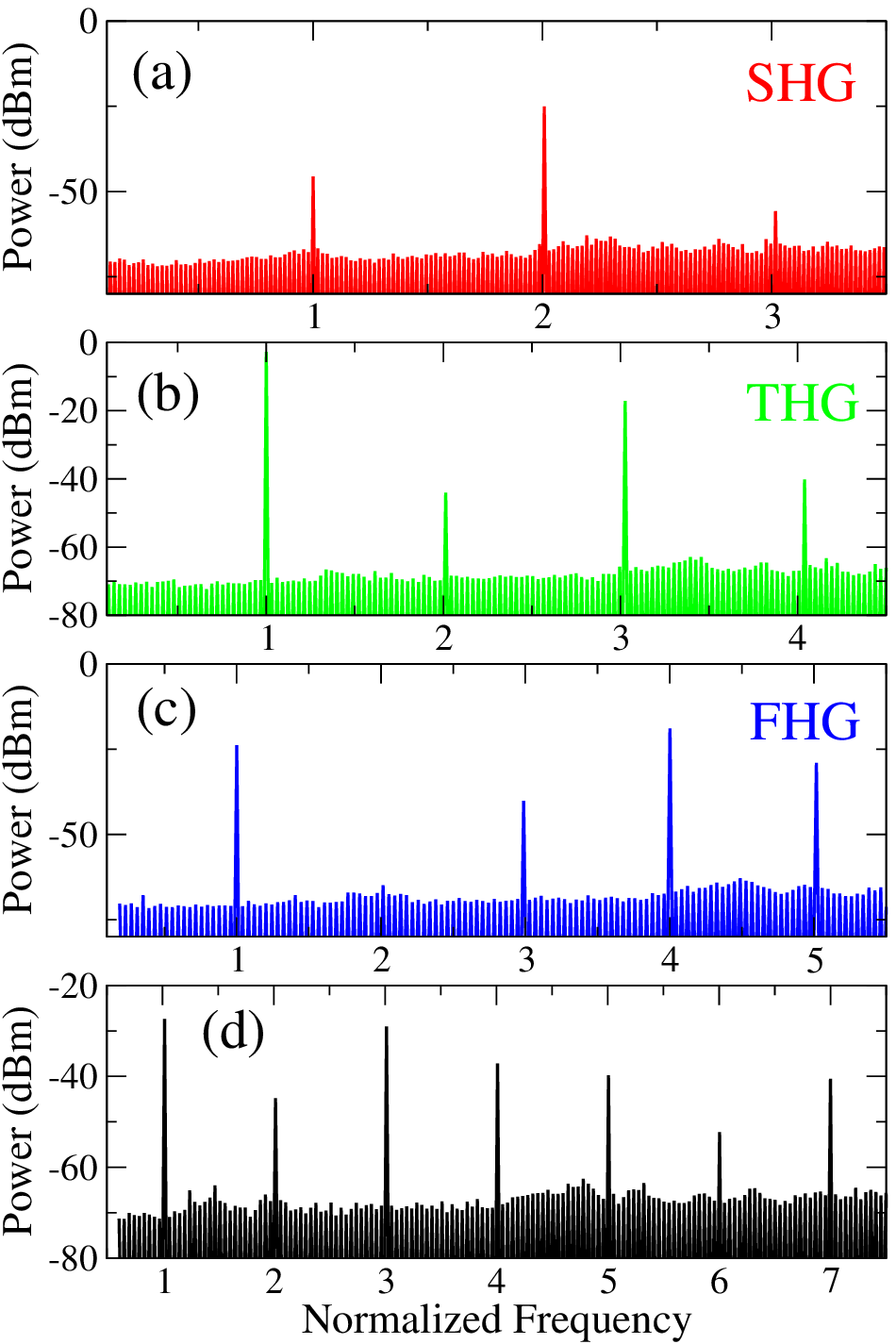}{harmonics}{(color online) Spectra of the signals
 detected by the receiving antenna located behind the nonlinear LHM
 slab for different source frequencies: (a) 3.415 GHz; (b) 2.29 GHz;
 (c) 1.733 GHz, (d) 1.668 GHz. Power at the input antenna is +30 dBm.
 }

In addition to the resonance shift measurements, we have studied
harmonic generation in nonlinear left-handed metamaterial. In
particular, we have measured the spectrum of the transmitted signal
for different frequencies of the incident electromagnetic wave. For
this purpose, the input antenna was fed by the signal generated by
an Agilent E4428C ESG vector signal generator and amplified by 38dB
amplifier. The signal detected by the receiving antenna behind the
metamaterial slab was analyzed using Agilent E4448A PSA Series
spectrum analyzer. Location of the antennas was the same as in the
previously described transmission coefficient measurements.

Figure~\rpict{harmonics} shows spectra of the signal detected by the
receiving antenna behind the nonlinear LHM slab. Varying the input
frequency, we observed efficient selective harmonic generation.
Namely second [Fig.~\rpict{harmonics}(a)], third
[Fig.~\rpict{harmonics}(b)] and fourth [Fig.~\rpict{harmonics}(c)]
harmonics were selectively generated. Moreover, the generation of a
comb-like signal was also observed [Fig.~\rpict{harmonics}(d)].

Selective generation of higher harmonics observed in our experiments
is related to the transmission properties of the metamaterial. A
particular harmonics dominates over fundamental harmonic and the
other higher harmonics when its frequency corresponds to the
transparency band. Results of the transmission coefficient
measurements performed on our nonlinear LHMs indicate a RH
transparency band with maximum transparency at around 7 GHz. This
value agrees well with the values of the higher harmonics dominating
in our measurements. Furthermore, presence of very high order
harmonics in the spectrum of the transmitted signal manifests strong
nonlinearity inside the metamaterial which potentially may lead to
significant enhancement of nonlinear processes in artificial
metamaterials as compared to conventional materials.

In conclusion, we have fabricated and analyzed the first tunable
nonlinear metamaterial operating at microwave frequencies. The
microwave metamaterial is composed of split-ring resonators and
wires where each split-ring resonator has a varactor diode, and it
can be tuned dynamically by varying the input power. We have shown 
that such nonlinear metamaterials demonstrate the power-dependent transmission and selective generation of higher harmonics.  We believe our experimental results open a door for a systematic study of 
many intriguing features of nonlinear metamaterials earlier considered only
theoretically for simplified models. 

Two of the authors (IS and YK) thank D. Powell, V. Shalaev, D.
Smith, C. Soukoulis, and A. Zharov for useful discussions and
suggestions. The work has been supported by the Australian Research
Council through the Discovery projects, by the Australian Academy of
Science through a travel grant, and by the Air Force Office of
Scientific Research (AFOSR) through the MURI program (grant
F49620-03-1-0420).


\end{sloppy}
\end{document}